\documentclass[a4paper,fleqn,usenatbib,referee]{mnras2}
\usepackage{newtxtext,newtxmath}

\usepackage[T1]{fontenc}

\DeclareRobustCommand{\VAN}[3]{#2}
\let\VANthebibliography\thebibliography
\def\thebibliography{\DeclareRobustCommand{\VAN}[3]{##3}\VANthebibliography}

\usepackage{graphicx}	
\usepackage{amsmath}	
\usepackage{float}
\usepackage{subfigure}
\usepackage{array}
\usepackage{enumitem}

\title[ULXs with He star companions]
{Ultraluminous X-ray sources with He star companions}

\author[L. Li et al.]{
        Luhan Li,$^{\rm 1,2,3}$\thanks{E-mail:liluhan@ynao.ac.cn}
        Bo Wang,$^{\rm 1,2}$\thanks{E-mail:wangbo@ynao.ac.cn} 
        Dongdong Liu,$^{\rm 1,2}$\thanks{E-mail:liudongdong@ynao.ac.cn}
	Yunlang Guo,$^{\rm 4,5}$
        Wen-Cong Chen,$^{\rm 6}$ and
        \newauthor
        Zhanwen Han$^{\rm 1,2}$
	\\
$^{1}$Yunnan Observatories, Chinese Academy of Sciences, Kunming 650216, China\\
$^{2}$International Centre of Supernovae, Yunnan Key Laboratory, Kunming 650216, China\\
$^{3}$University of Chinese Academy of Sciences, Beijing 100049, China\\
$^{4}$School of Astronomy and Space Science, Nanjing University, Nanjing 210023, China\\
$^{5}$Key Laboratory of Modern Astronomy and Astrophysics, Nanjing University, Ministry of Education, Nanjing 210023, China\\
$^{6}$School of Science, Qingdao University of Technology, Qingdao 266525, China
}

\date{Accepted XXX. Received YYY; in original form ZZZ}

\pubyear{2024}

\begin{document}
\label{firstpage}
\pagerange{\pageref{firstpage}--\pageref{lastpage}}
\maketitle

\begin{abstract}
Ultraluminous X-ray sources (ULXs) are non-nuclear point-like
objects observed with extremely high X-ray luminosity that exceeds the Eddington limit of a $\rm10\,M_\odot$ black hole.
A fraction of ULXs has been confirmed to contain neutron star (NS) accretors due to the discovery of their X-ray pulsations.
The donors detected in NS ULXs are usually luminous massive stars because of the observational biases.
Recently, the He donor star in NGC 247 ULX-1 has been identified, which is the first evidence of a He donor star in ULXs.
In this paper, we employed the stellar evolution code MESA to investigate the formation of ULXs through the NS+He star channel, in which a He star transfers its He-rich material onto the surface of a NS via Roche-lobe overflow. 
We evolved a large number of NS+He star systems and provided the parameter space for the production of ULXs.
We found that the initial NS+He star systems should have $\rm\sim 0.7-2.6 \, M_\odot$ He star and $\rm \sim 0.1-2500\, d$ orbital period for producing ULXs, eventually evolving into intermediate-mass binary pulsars.
According to binary population synthesis calculations, we estimated that the Galactic rate of NS ULXs with He donor stars is in the range of $\sim1.6-4.0\times10^{-4}\,{\rm yr}^{-1}$, and that there exist $\sim7-20$ detectable NS ULXs with He donor stars in the Galaxy.
\end{abstract}




\section{INTRODUCTION}

Ultraluminous X-ray sources (ULXs) are non-nuclear point-like sources found in external galaxies \citep[e.g.][]{2011NewAR..55..166F,2017ARA&A..55..303K}.
Under the assumption of isotropic emission, their X-ray luminosity exceeds $10^{39}\, \rm{erg\, s^{-1}}$, which is about the Eddington limit of a $10\, \rm M_\odot$ black hole (BH).
Most of the ULXs are believed to be X-ray binaries (XRB) powered by accretion onto BH or neutron star (NS) in unusual and short-lived stages \citep[see][]{2001ApJ...552L.109K}.
The stage of ULXs may play an important role in the formation of close systems with two compact objects, which are the potential targets of gravitational wave telescopes \citep{2017A&A...604A..55M,2021AstBu..76....6F}.

Up to now, there are two main evolutionary ways to explain the observed ULXs as follows:
(1) Intermediate-mass BH (IMBH; $\rm 10^2 - 10^5\, M_{\odot}$) with sub-Eddington accretion rate is a proposed way to explain the ULXs with the peak luminosities $\ga 10^{41}\, \rm{erg\ s^{-1}}$ \citep[see e.g.][]{1999ApJ...519...89C,2007Natur.445..183M}.
ESO 243-49 HLX-1 was considered to be a strong IMBH candidate  having a $\rm 6-200\,\times10^3\,M_\odot $
accretor \citep{2014A&A...569A.116S}.
Recently, \cite{2023ApJ...956....3S} suggested that the ULX CXO J133815.6+043255 is also a candidate of IMBH residing in the outskirts of NGC 5252.
(2) The other evolutionary way is stellar-mass compact objects (BHs or NSs) with super-Eddington accretion rate \citep[e.g.][]{2001ApJ...552L.109K,2002ApJ...568L..97B,2007MNRAS.377.1187P,2009MNRAS.397.1836G,2018ApJ...856..128W}.
It has been suggested that M101 ULX-1 consists of a $\rm \sim20-30\,M_\odot$ BH and a Wolf-Rayet star with an orbital period of $\rm 8.2\,d$ \citep{2013Natur.503..500L}.
M82 X-2 is the first confirmed ULX with an accreting NS due to the discovery of a coherent periodicity of $1.37\, \rm s$ \citep{2014Natur.514..202B}. 
\cite{2024arXiv240313973E} suggest that there may be a NS accretor in NGC 4190 ULX-1 based on the spectral and timing analysis of a broadband NICER+NuSTAR observation.
Binary population synthesis (BPS) calculations indicate that NS accretors dominate the ULX populations \citep[e.g.][]{2015ApJ...802L...5F,2015ApJ...802..131S,2017ApJ...846...17W,2019ApJ...886..118S,2024A&A...682A..69M}.

Up to date, only about 10 NS ULXs have been confirmed \citep{2024A&A...682A..69M}.
However, the nature of the donors in NS ULXs is still not clear. 
The donor in M51 ULX-7 was suggested to be an OB giant star with mass $\rm\ga 8\, M_{\odot}$ \citep{2020ApJ...895...60R}.
\cite{2019ApJ...883L..34H} discovered a red supergiant star in NGC 7793 P-13, suggesting that the donor mass range is $\rm8 - 10\,M_{\odot}$.  
SMC X-3 is also a NS ULX containing a Be star with mass $\rm\ga 3.7\, M_{\odot}$ \citep{2017ApJ...843...69W,2017A&A...605A..39T}.
In general, the donors detected in NS ULXs are usually luminous massive stars due to the observational biases favoring bright stars.

Recently, a He donor star in NGC 247 ULX-1 has been identified as no Balmer lines can be seen in the spectrum from the optical counterpart \citep{2023ApJ...947...52Z}.
This is the first evidence for the presence of a He star in ULXs, 
providing a support for the NS+He channel and
indicating that the NS+He star channel is a possible evolutionary way for the production of ULXs.
\cite{2019ApJ...886..118S} suggested that NS+He star binaries can significantly contribute to the ULX population by using the population synthesis study.

An important way to explain the super-Eddington rate is the formation of an accretion disk that receives material at a super-Eddington rate \citep[super-critical disk model; see e.g.][]{1973A&A....24..337S,1999AstL...25..508L,2007MNRAS.377.1187P,2015MNRAS.447.3243M}.
In this model, the structure of the accretion disc is not a thin disk but a geometrically thick disk \citep[see][]{2017ARA&A..55..303K}, in which the Eddington limit is maintained at any position of the accretion disc.
To date, it has been suggested that the super-Eddington luminosity observed in ULXs can be well explained by adopting the super-critical disk model \citep[see e.g.][]{2016MNRAS.458L..10K,2020MNRAS.494.3611K,2015ApJ...802..131S,2017MNRAS.470L..69M,2017ApJ...846...17W,2019ApJ...886..118S,2023MNRAS.526.2506L}.

It is worth noting that \cite{2019A&A...626A..18C} proposed a model for a super-Eddington accretion disc , considering the advection of heat and mass loss by the wind around magnetized NSs.
In the observations, by using the broad energy band of Insight-HXMT, \cite{2020MNRAS.491.1857D} discovered a sharp state transition of the timing and spectral properties of Swift J0243.6+6124 at super-Eddington accretion rate, suggesting that this source has a magnetized NS accretor. 
The cyclotron resonance scattering features have been detected in Swift J0243.6+6124 \citep{2022ApJ...933L...3K} and RX J0209.6-7427 \citep{2022ApJ...938..149H}, which supports the multi-pole magnetic fields in the model of \cite{2019A&A...626A..18C}.
The magnetar accretor model in ULXs is still under debate \citep[e.g.][]{2016A&AT...29..183P,2019MNRAS.485.3588K,2023MNRAS.526.2506L}.

In this article, we aim to investigate the formation of ULXs through the NS+He star channel using the super-critical disk model systematically and consider the future evolution of ULXs from this channel.
In Section~\ref{sec:2}, we introduce the numerical methods for the evolution of NS+He star systems and give the corresponding results in Section~\ref{sec:3}.
In Section~\ref{sec:4}, we present the BPS methods and the corresponding results. 
Finally, we make a discussion in Section~\ref{sec:5} and a brief summary in Section~\ref{sec:6}.

\section{NUMERICAL METHODS for binary evolution}
\label{sec:2}
\subsection{Basic settings for binary evolution}
We carried out detailed binary evolution calculations of NS+He star systems that can undergo the super-Eddington rate stage by using the stellar evolution code Modules for Experiments in Stellar Astrophysics 
\citep[MESA, version number 15140;][]{2011ApJS..192....3P,2013ApJS..208....4P,2015ApJS..220...15P,2018ApJS..234...34P,2019ApJS..243...10P}. 
In our simulations, a typical Population I metallicity (Z = 0.02) is adopted, and the zero-age He main-sequence star models are composed of $98\%$ helium and $2\%$ metallicity.
The NSs are treated as point masses and the initial NS masses are $\rm1.4\,M_\odot$.
We vary the He star masses from 0.5 to $\rm3.0\,M_\odot$ by steps of $\rm0.1\,M_\odot$ and the binary orbital periods (in units of days) logarithmically from $-2.0$ to 4.0 by steps of 0.1.
We considered the orbital angular momentum loss due to the mass loss and the gravitational wave radiation.
We did not consider the tide effect in binary evolution.
Magnetic braking is not considered, because it is usually used for Sun-like stars with radiative cores and convective envelopes \citep[e.g.][]{1983ApJ...275..713R,2013ApJ...775...27C,2015ApJS..220...15P,2024MNRAS.527.7394G}.

During the binary evolution, the He star fills its Roche-lobe and transfers He-rich matter to NS via Roche-lobe overflow (RLOF) when it evolves to the He subgiant phase.
We adopt the Ritter mass-transfer scheme to calculate the mass-transfer rate \citep[see][]{1988A&A...202...93R}.
For the NSs, we set the mass-transfer efficiency with $\alpha = 0$, $\beta = 0.7$ \citep[see][]{2021MNRAS.503.3540C,2021ApJ...922..158L} and $\delta = 0$ in which $\alpha$, $\beta$, and $\delta$ are the fractions of the mass loss from the vicinity of the donor star, the vicinity of the NS, and the circumbinary co-planar toroid, respectively \citep{2006csxs.book..623T}.
Accordingly, the NS mass accretion rate ($\dot{M}_{\rm acc}$) is 
equal to $(1-\beta)\dot{M}_{\rm tr}$, where $\dot{M}_{\rm tr}$ is the mass-transfer rate. 
If the NS mass accretion rate is larger than the Eddington accretion rate of He accretion, we assume that the excess matter is ejected from the vicinity of the NSs, taking away the specific orbital angular momentum of the accreting NSs \citep[see][]{2020ApJ...900L...8C,2021MNRAS.506.4654W}.

It is difficult to observe the NS ULXs with too short mass-transfer timescale.
In our simulation, the NS$+$He star systems are assumed to be ULXs if the timescale for $\dot{M}_{\mathrm{tr}}$>$\dot{M}_{\mathrm{Edd}}$ is larger than $\rm 0.1\,Myr$.
It is worth noting that the maximum ULX lifetime for NS ULXs with low luminosities is $\rm\sim 1.0\,Myr$ \citep[see][]{2020A&A...642A.174M}.

\subsection{X-ray luminosity calculation}
The X-ray accretion luminosity can be simply estimated by using
\begin{equation}
L_{\mathrm{acc}}=\eta \dot{M}_{\mathrm{acc}} {c}^2,
\label{equ:Lacc}
\end{equation}
where $\eta$ is the radiative efficiency of the accretion flow, and $c$ is the speed of the light in vacuum.
For the NS accretors, the radiative efficiency $\eta \approx 0.10-0.20 $ and we set $\eta = 0.15$ in this work \citep[see e.g.][]{2017ARA&A..55..303K,2023pbse.book.....T}.
Accordingly, the Eddington accretion rate of He accretion is $\rm\sim 4\times 10^{-8}\, M_\odot \,yr^{-1}$ and the Eddington luminosity $L_{\rm Edd}$ is $\rm\sim 3.5\times10^{38}\,erg\, s^{-1}$.

When the  $\dot{M}_{\mathrm{tr}}$ exceeds the $\dot{M}_{\mathrm{Edd}}$, the accretion disk becomes geometrically thick.
A strong stellar wind appears on the surface of the disk, taking away excess material and angular momentum, thereby maintaining a local Eddington limit on the disk.
We adopt the accretion disk model of \cite{1973A&A....24..337S} to calculate the accretion luminosity, which can be written as 
\begin{equation}
L_{\mathrm{X}} \simeq L_{\mathrm{Edd}}(1+\ln \dot{m}),
\label{equ:Lx}
\end{equation}
where $\dot{m}$ is the ratio of the mass-transfer rate to the Eddington accretion rate.
For simplicity, we adopt the fixed Eddington limit ($\rm 4\times 10^{-8}\, M_\odot \,yr^{-1}$) to calculate the X-ray luminosity during the mass-transfer process, because the amount of mass accreted onto NSs does not change the Eddington limit significantly.

Additionally, the stellar wind on the surface of the disc forms a hollow cone, and the emitted X-rays escape through this cone, which results in an additional increase in the luminosity when observers are viewing the source face-to-face \citep[e.g.][]{2001ApJ...552L.109K,2009MNRAS.393L..41K}.
By considering the geometrical beaming effect of \cite{2001ApJ...552L.109K} and \cite{2009MNRAS.393L..41K}, we can only see the emitted X-ray in directions within one of the radiation cones. 
The isotropic-equivalent observed X-ray luminosity is defined as follows:
\begin{equation}
L_{\mathrm{X}} \simeq \frac{L_{\mathrm{Edd}}}{b}(1+\ln \dot{m}),
\label{equ:Lxb}
\end{equation}
where $b$ is the beaming factor reflecting the geometrical collimation of the emission from the thick disc. 
\cite{2009MNRAS.393L..41K} gave an approximate formula of $b$ as follows:
\begin{equation}
b= \begin{cases}\frac{73}{\dot{m}^2}, & \text { if } \dot{m}>8.5, \\ 1, & \text { otherwise. }\end{cases}
\end{equation}
Note that the beaming effect leads to an increase in the isotropic X-ray luminosity of ULXs, but the probability of detecting ULXs along the beam is reduced.

\section{BINARY EVOLUTION RESULTS}
\label{sec:3}
In order to investigate the formation of NS ULXs with He star donors, we performed a large number of NS+He star binary evolution computations, and thus we obtained a dense grid of binaries.
In Table~\ref{tab:evolutionary properties}, we show the main evolutionary features of some selected NS+He star systems that can form NS ULXs.
In this table, we explored the effect of different initial orbital periods (see sets 1-5) and initial He star masses (see sets 6-10) on the final results.

\begin{table*}
	\centering
 \renewcommand{\arraystretch}{1.2}
	\caption{The evolutionary properties of NS+He star systems with different initial He star masses and initial orbital periods, in which the initial mass of NSs is assumed to be $\rm1.4\,M_\odot$. $M_2^{\rm i}$ and $P_{\rm orb}^{\rm i}$ are the initial He star mass in solar masses and the initial orbital period in days; $M_2^{\rm f}$, $P_{\rm orb}^{\rm f}$ and $\Delta M_{\rm NS}$ are the final He star mass, the final orbital period and the NS mass increase; $t_{\rm ULX}$ and $L_{\rm x,max}$ are the duration of ULX lifetime and the brightest X-ray luminosity based on Equation~\ref{equ:Lxb}. Note that the final orbital period ($P_{\rm orb}^{\rm f}$) here represents the binary period at the end of mass-transfer process.}
          \label{tab:evolutionary properties}
	\begin{tabular}{ l  c c c ccc c c c  l }
		\hline 
		Set	&$M_{\rm 
  2}^{\rm i}$ & log$\,P_{\rm orb}^{\rm i}$ & $M_{\rm 2}^{\rm f}$ & ${\rm log}\, P_{\rm orb}^{\rm f}$ & $\Delta\,M_{\rm NS}$ & $t_{\rm ULX}$ & ${\rm log}\, L_{\rm x,max}$ \\
		& ($\rm M_\odot$) & (d) & ($\rm M_\odot$) &	 (d) & ($\rm M_\odot$) & (Myr) & (erg/s)\\
		\hline 
		01	& $1.0$ & $-1.0$ &  $0.76$ & $-0.82$ &$0.0224$ & $0.69$ & $39.33$\\
            02   & $1.0 $ & $\ \ \ 0.0 $ & $0.83$ & $\ \ \ 0.15$ & $0.0065 $ & $0.19$ & $40.63$\\
            03   & $1.0 $ & $\ \ \ 1.0 $ & $0.84$ & $\ \ \ 1.14$ & $0.0046 $ & $0.14$ & $41.08$\\
            04   & $1.0 $ & $\ \ \ 2.0 $ & $0.86$ & $\ \ \ 2.12$ & $0.0036 $ & $0.11$ & $41.21$\\
            05   & $1.0 $ & $\ \ \ 3.0 $ & $0.87$ & $\ \ \ 3.10$ & $0.0033 $ & $0.10$ & $41.38$\\
            \\
            06   & $1.2 $ & $-1.0$ & $0.82$ & $-0.75$ & $0.0170$ & $0.52$ & $40.06$\\
            07   & $1.4 $ & $-1.0$ & $0.88$ & $-0.75$ & $0.0129$ & $0.39$ & $40.65$\\
            08   & $1.6 $ & $-1.0$ & $0.94$ & $-0.72$ & $0.0094$ & $0.28$ & $41.17$\\
            09   & $1.8 $ & $-1.0$ & $1.01$ & $-0.75$ & $0.0068$ & $0.20$ & $41.83$\\
            10  & $2.0 $ & $-1.0$ & $1.08$ & $-0.80$ & $0.0053$ & $0.16$ & $42.33$\\
		\hline 
	\end{tabular}
\end{table*}

\begin{figure*}
\centering
\includegraphics[width=1.0\textwidth]{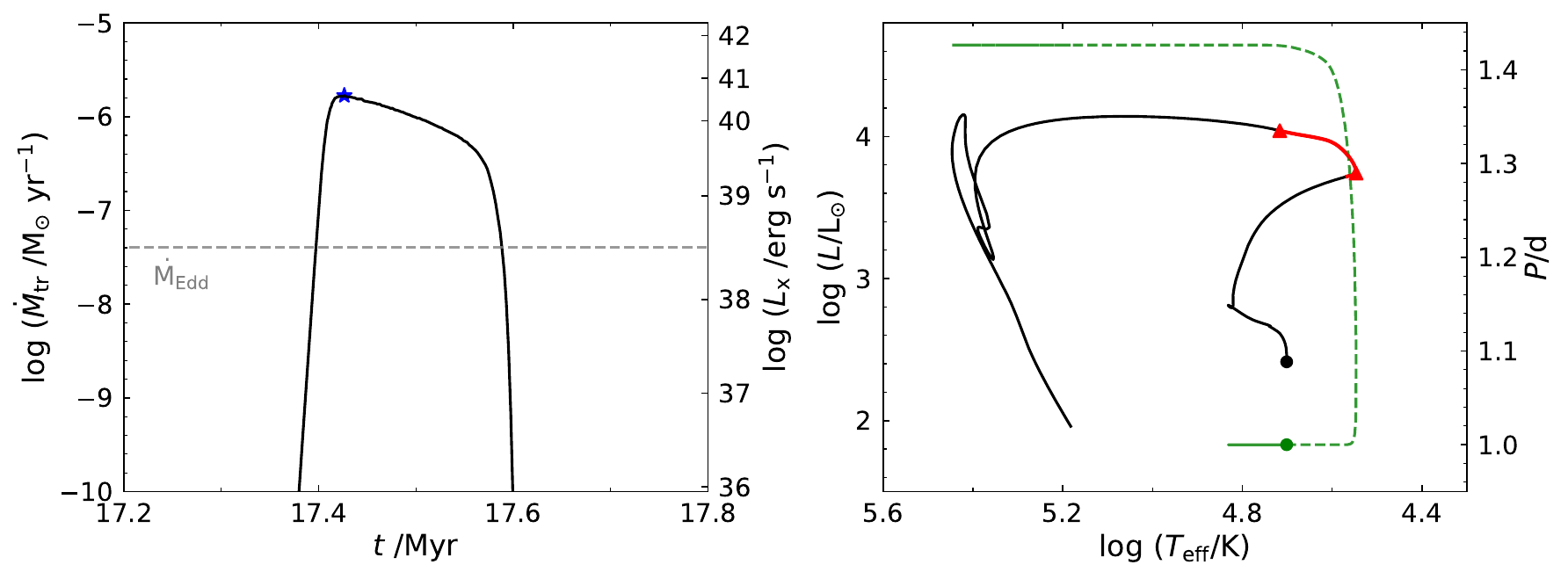}
\caption{A representative example for the evolution of a NS+He star system that can form a ULX, in which the initial NS mass is set to be $\rm1.4\,M_\odot$.
The left panel shows the evolution of $\dot{M}_{\rm tr}$ as a function of time for the binary evolution calculations. 
The grey dashed line stands for the critical Eddington accretion rate. 
The blue asterisks stand for the maximum mass-transfer rate and the maximum X-ray luminosity ($\sim 4.0\times10^{40}\, \rm erg\,s^{-1}$) due to Equation ~\ref{equ:Lxb}.
In the right panel, the black solid line stands for the evolution track of the He star in the H-R diagram, and the green dashed line shows the evolution of the orbital period with the effective temperature.
The round dots stand for the start of the binary evolution.
The red thick line corresponds to the super-Eddington rate stage.}
   \label{fig:sample}
\end{figure*}

\begin{figure}
\centering
\includegraphics[width=0.6\columnwidth]{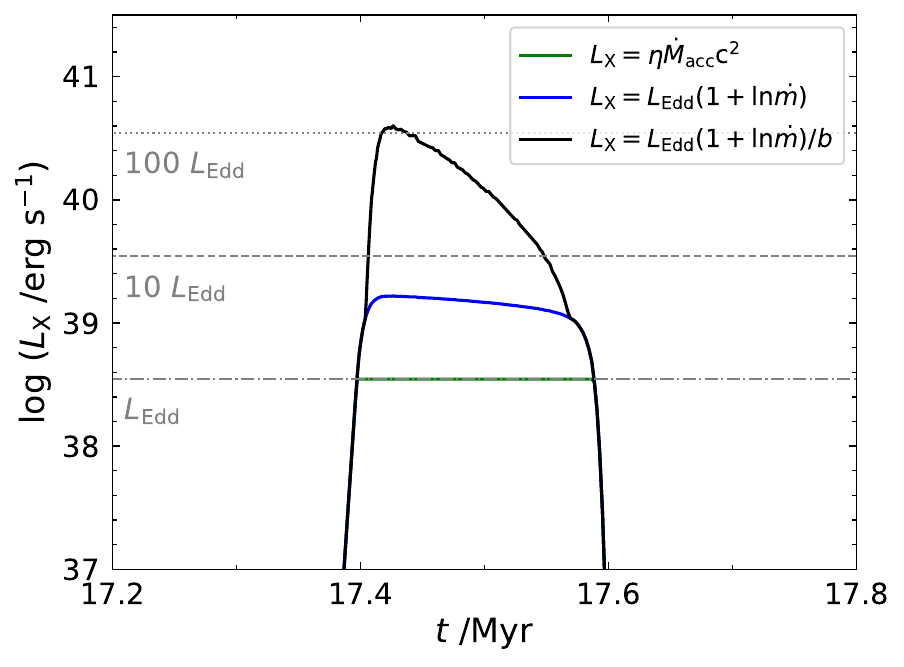}
\caption{The X-ray luminosity comparison by using different models in the sample of Figure~\ref{fig:sample}. The green, blue and black curves represent the NS accretion luminosity shown in Equation~\ref{equ:Lacc}, the X-ray luminosity of critical disk shown in Equation~\ref{equ:Lx}, and isotropic-equivalent observed X-ray luminosity shown in Equation~\ref{equ:Lxb}.}
   \label{fig:Lx}
\end{figure}

\begin{figure*}
\centering
\includegraphics[width=0.85\textwidth]{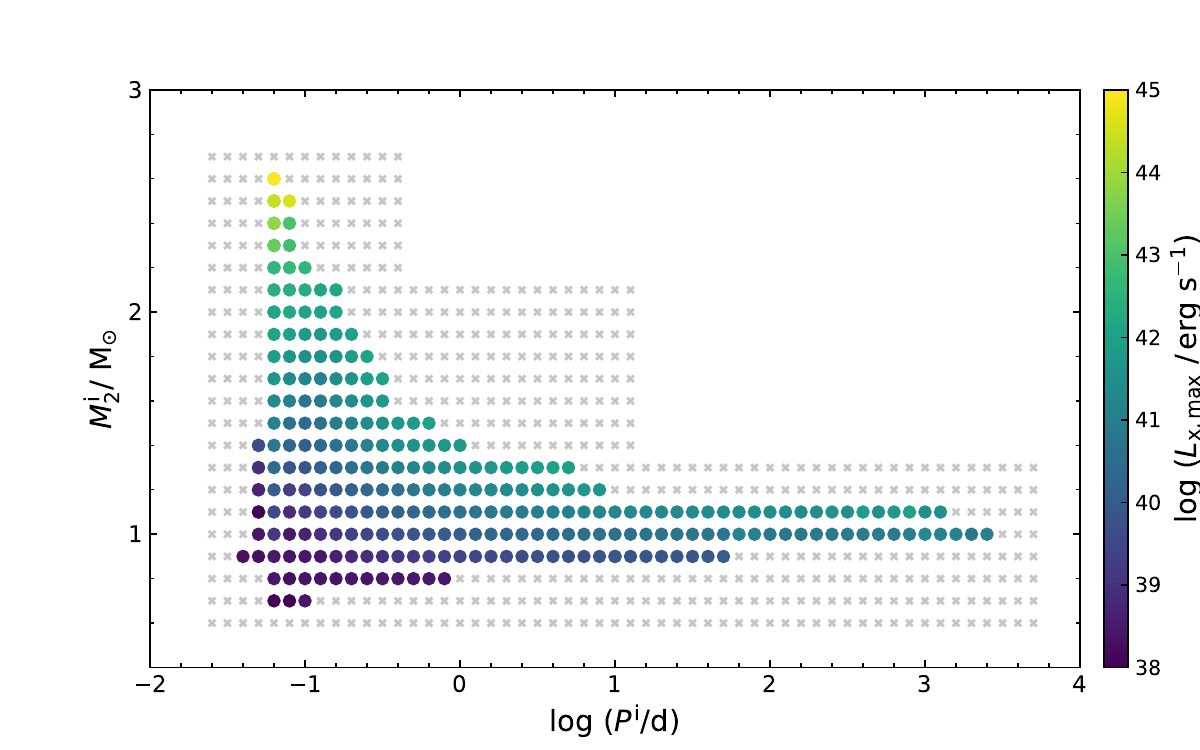}
\caption{Initial parameter space of NS+He star systems that will form ULXs in the $\mathrm{log}\,P^{\rm i}-M_2^{\rm i} $ plane, in which we set $M_{\rm NS}^{\rm i}=1.4\,\mathrm{M}_\odot$. The colored circles denote systems that will experience the super-Eddington rate stage and then show as ULXs during the mass-transfer process. The grey crosses indicate systems that will not form ULXs. The colorbar represents the maximum X-ray luminosity based on Equation ~\ref{equ:Lxb}. }
   \label{fig:ps}
\end{figure*}

\begin{figure*}
\centering
\includegraphics[width=0.7\textwidth]{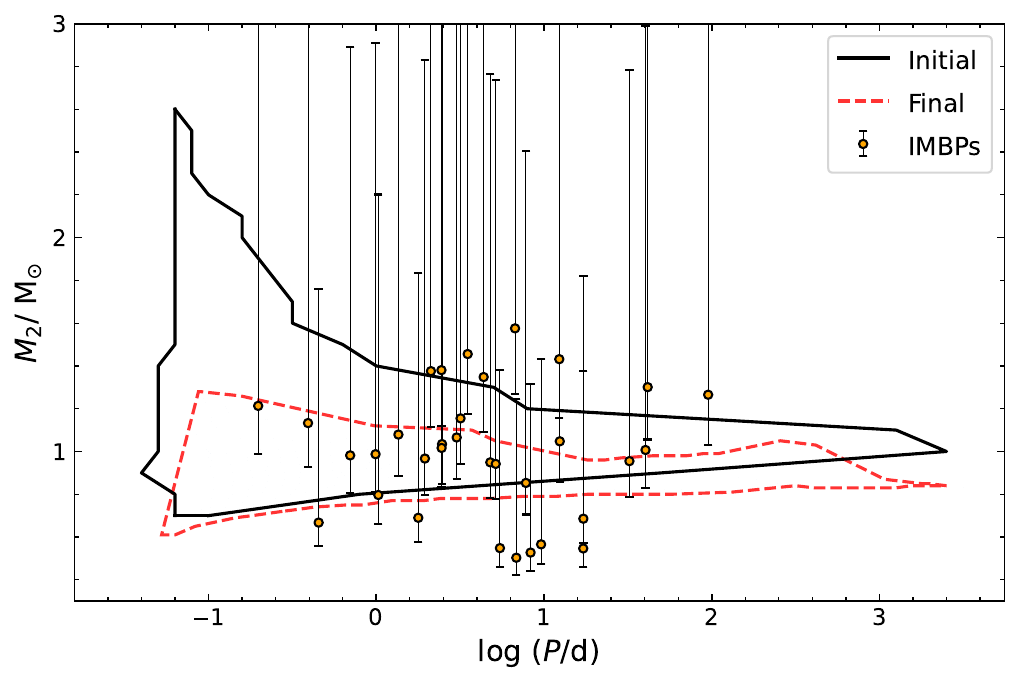}
\caption{Comparison between the parameter space and the observed IMBPs in the $\mathrm{log}\,P-M_2$ plane. The solid black and dashed red contours represent the initial and final parameter space, respectively. The error bars stand for the 33 observed IMBPs listed in Table~\ref{tab:IMBPs}.}
   \label{fig:IMBP}
\end{figure*}

\begin{table}
\renewcommand{\arraystretch}{1.4}
\begin{center}
 \caption{The parameters of 33 observed IMBPs on the galactic disk, which are taken from the ATNF Pulsar Catalogue in 2023 November \citep[][see http://www.atnf.csiro.au/research/pulsar/psrcat]{2005AJ....129.1993M}. 
 The median masses of the WD companions in IMBPs are computed by supposing a typical pulsar mass of $\rm1.35\,M_\odot$ and an orbital inclination angle of $60^{\circ}$. The lower limit of companion mass marks an inclination angle of $90^{\circ}$, whereas the upper limit represents a $90\%$ confidence probability limit by assuming the orbital inclination angle is $26^{\circ}$. }
\begin{tabular}{ccccccccc}
\hline \hline
 $\rm No.$ & $\rm Pulsars$ & $\rm P_{\rm spin}/ms$ & $\rm P_{\rm orb}/d$ & $\rm M_{\rm WD}/M_{\odot}$\\
\hline
$1$ & $\rm  J0621+1002$ & $28.85$ & $8.32$ & $ 0.53_{\texttt{-}0.09}^{\texttt{+}0.79} $\\ 
$2$ & $\rm  J0700+6418$ & $195.67$ & $1.03$ & $ 0.80_{\texttt{-}0.14}^{+1.40} $\\ 
$3$ & $\rm  J0709+0458$ & $34.43$ & $4.37$ & $ 1.35_{\texttt{-}0.26}^{+3.07} $\\ 
$4$ & $\rm  J0721-2038$ & $15.54$ & $5.46$ & $ 0.55_{\texttt{-}0.09}^{+0.83} $\\ 
$5$ & $\rm  J1022+1001$ & $16.45$ & $7.81$ & $ 0.85_{\texttt{-}0.14}^{+0.55} $\\ 
$6$ & $\rm  J1101-6424$ & $5.11$ & $9.61$ & $ 0.57_{\texttt{-}0.10}^{+0.86} $\\ 
$7$ & $\rm  J1141-6545$ & $393.9$ & $0.2$ & $ 1.21_{\texttt{-}0.22}^{+2.62} $\\ 
$8$ & $\rm  J1157-5112$ & $43.59$ & $3.51$ & $ 1.46_{\texttt{-}0.28}^{+3.45} $\\ 
$9$ & $\rm  J1227-6208$ & $34.53$ & $6.72$ & $ 1.58_{\texttt{-}0.31}^{+3.90} $\\ 
$10$ & $\rm  J1244-6359$ & $147.27$ & $17.17$ & $ 0.69_{\texttt{-}0.12}^{+1.13} $\\ 
$11$ & $\rm  J1337-6423$ & $9.42$ & $4.79$ & $ 0.95_{\texttt{-}0.17}^{+1.81} $\\ 
$12$ & $\rm  J1435-6100$ & $9.35$ & $1.35$ & $ 1.08_{\texttt{-}0.20}^{+2.19} $\\ 
$13$ & $\rm  J1439-5501$ & $28.63$ & $2.12$ & $ 1.38_{\texttt{-}0.27}^{+3.17} $\\ 
$14$ & $\rm  J1454-5846$ & $45.25$ & $12.42$ & $ 1.05_{\texttt{-}0.19}^{+2.10} $\\ 
$15$ & $\rm  J1525-5545$ & $11.36$ & $0.99$ & $ 0.99_{\texttt{-}0.18}^{+1.92} $\\ 
$16$ & $\rm  J1528-3146$ & $60.82$ & $3.18$ & $ 1.15_{\texttt{-}0.21}^{+2.44} $\\  
$17$ & $\rm  J1618-4624$ & $5.93$ & $1.78$ & $ 0.69_{\texttt{-}0.12}^{+1.14} $\\ 
$18$ & $\rm  J1658+3630$ & $33.03$ & $3.02$ & $ 1.07_{\texttt{-}0.20}^{+2.15} $\\ 
$19$ & $\rm  J1727-2946$ & $27.08$ & $40.31$ & $ 1.01_{\texttt{-}0.18}^{+1.98} $\\ 
$20$ & $\rm  J1750-2536$ & $34.75$ & $17.14$ & $ 0.55_{\texttt{-}0.09}^{+0.83} $\\ 
$21$ & $\rm  J1757-5322$ & $8.87$ & $0.45$ & $ 0.67_{\texttt{-}0.11}^{+1.09} $\\ 
$22$ & $\rm  J1802-2124$ & $12.65$ & $0.7$ & $ 0.98_{\texttt{-}0.17}^{+1.91} $\\ 
$23$ & $\rm  J1932+1756$ & $41.83$ & $41.51$ & $ 1.30_{\texttt{-}0.24}^{+2.91} $\\ 
$24$ & $\rm  J1933+1726$ & $21.51$ & $5.15$ & $ 0.94_{\texttt{-}0.16}^{+1.80} $\\  
$25$ & $\rm  J1938+6604$ & $22.26$ & $2.47$ & $ 1.03_{\texttt{-}0.18}^{+2.07} $\\ 
$26$ & $\rm  J1949+3106$ & $13.14$ & $1.95$ & $ 0.97_{\texttt{-}0.17}^{+1.86} $\\ 
$27$ & $\rm  J1952+2630$ & $20.73$ & $0.39$ & $ 1.13_{\texttt{-}0.20}^{+2.36} $\\ 
$28$ & $\rm  J2045+3633$ & $31.68$ & $32.3$ & $ 0.95_{\texttt{-}0.16}^{+1.84} $\\ 
$29$ & $\rm  J2053+4650$ & $12.59$ & $2.45$ & $ 1.02_{\texttt{-}0.19}^{+2.01} $\\ 
$30$ & $\rm  J2145-0750$ & $16.05$ & $6.84$ & $ 0.50_{\texttt{-}0.08}^{+0.74} $\\ 
$31$ & $\rm  J2222-0137$ & $32.82$ & $2.45$ & $ 1.38_{\texttt{-}0.26}^{+3.19} $\\ 
$32$ & $\rm  J2305+4707$ & $1066.37$ & $12.34$ & $ 1.43_{\texttt{-}0.27}^{+3.37} $\\ 
$33$ & $\rm  J2338+4818$ & $118.71$ & $95.26$ & $ 1.27_{\texttt{-}0.24}^{+2.79} $\\ 
\hline
\label{tab:IMBPs}
\end{tabular}

\end{center}
\end{table}

\subsection{A representative example for binary evolution}

Figure~\ref{fig:sample} presents a typical example (see set 02 in Table~\ref{tab:evolutionary properties}) of binary evolution computations with initial parameters of $(M_{\rm NS}^{\rm i}/\mathrm{M}_{\odot}, M_{\rm 2}^{\rm i}/\mathrm{M}_\odot, P_{\rm orb}^{\rm i}/{\rm d}) = (1.4, 1.0, 1.0)$, in which $M_{\rm NS}^{\rm i}$, $M_{\rm 2}^{\rm i}$ and $P_{\rm orb}^{\rm i}$ is the initial mass of NS, the initial mass of He donor, and the initial orbital period, respectively. 
The He star experiences the He-core burning for about $\rm 16.1\,Myr$.
After the exhaustion of the central He core, the envelope of the He star expands rapidly, and the He star fills its Roche-lobe after about $\rm 1.2\,Myr$. 
The mass-transfer rate exceeds the Eddington rate quickly and will last about $\rm0.19\,Myr$, during which the binary system shows as a ULX with maximum X-ray luminosity of about $ 4\times10^{40}\, \rm{erg\ s^{-1}}$ based on Equation~\ref{equ:Lxb}.
Due to the rapid mass-transfer, the $1.0\,\mathrm{M}_\odot$ He star eventually evolves to be a $0.82\,\mathrm{M}_\odot$ CO WD and the binary orbital period increases to $\rm1.43\,d$.
After the mass-transfer process, the NS has accreted $\rm\sim 0.0065\,M_\odot$ material and shows as a pulsar \citep{2012MNRAS.425.1601T}. 
Eventually, this NS+He star system evolves into a pulasr+CO WD system, which is recognized as an IMBP in observations.
Note that the IMBPs are usually considered to consist of a pulsar and a CO/ONe WD with $M_{\rm wd}\ga 0.45\,\mathrm{M}_\odot $ \citep[see e.g.][]{1996ApJ...469..819C,2001ApJ...548L.187C,2001ApJ...547L..37E}.

Figure~\ref{fig:Lx} shows the comparison of the estimated X-ray luminosities by using different methods in the example shown in Figure~\ref{fig:sample}.
The green line represents the traditional accretion luminosity by using Equation~\ref{equ:Lacc}.
Thus, the maximum accretion luminosity is Eddington luminosity. 
The blue curve stands for the accretion luminosity by using the supercritical disk model based on Equation~\ref{equ:Lx}, which can reach several times of Eddington luminosity.
Due to the consideration of the beaming effect, the isotropic observed X-ray luminosity by using Equation~\ref{equ:Lxb} is shown in black curve and the maximum luminosity ($\rm 4.0\times10^{40}\,erg\,s^{-1}$) can exceed a hundred times Eddington luminosity.

\subsection{Initial parameter space of ULXs}
Figure~\ref{fig:ps} shows the initial parameter space of NS+He star systems that will experience super-Eddington rate stage to show as ULXs during the mass-transfer process. 
In order to form ULXs, the initial NS+He star systems should have $\rm\sim 0.7-2.6 \, M_\odot$ He star and $\rm \sim 0.1-2500\, d$ orbital period.  
The systems with longer orbital periods and larger initial donor masses will exhibit brighter X-ray luminosity due to the higher mass-transfer rate.
After the mass-transfer process, the He stars with initial masses less/greater than $\sim\rm 1.9\,M_\odot$ will evolve into CO/ONe WDs.
Eventually, these initial NS+He star binaries will evolve into IMBPs.

The grey crosses denote systems that will not form ULXs for different reasons as follows:
(1) The binaries beyond the upper boundary will stay in the super-Eddington stage for too short time due to excessive initial donor masses and mass-transfer rate.
(2) The lower boundary is constrained by the low mass-transfer rate when the donors fill their Roche-lobes.
(3) The initial orbital periods of binaries beyond the right boundary are so large that the mass-transfer rate is too high and ULX lifetime is less than $\rm 0.1\,Myr$.
(4) The left boundary is set by the condition that RLOF has been started when the He star is in the zero-age main-sequence (ZAMS) stage.

\subsection{Resulting IMBPs}

In our simulations, the NS+He star systems in the initial parameter space undergo the super-Eddington rate stage and show as ULXs during the mass-transfer process, the NSs accrete He-rich material and show as pulsars, and these systems eventually evolve into IMBPs.
The resulting IMBPs have $\rm\sim 0.05-3200\,d$ orbital period and $\rm \sim 0.6-1.3\,M_\odot$ WD donor mass.
In Figure~\ref{fig:IMBP}, we show the comparison between the parameter space and the 33 observed IMBPs that are shown in Table~\ref{tab:IMBPs}. 
The NS+WD binaries in the final parameter space have $\sim\rm 0.6-1.3\,M_\odot$ WD and $\sim \rm 0.2-3000\,d$ orbital period.
From this figure, the median parameter of the 14 IMBPs with $\rm\sim1.0\,M_\odot$ WD companions can be covered in our final parameter space.

Note that the more classical evolutionary way to form the IMBPs is the intermediate-mass X-ray binary (IMXB) channel, i.e., a NS accretes a $\rm 2-10\, M_\odot$ donor star \citep[e.g.][]{1975ApJ...198L.109V,2000ApJ...530L..93T}.
In this study, the NS+He star systems that could form ULXs can contribute to the part of the observed IMBPs. 
Although some NS+He star systems are not in our initial parameter space of ULXs, they may also eventually evolve into IMBPs \citep[see][]{2013MNRAS.432L..75C,2019MNRAS.490..752T}.


\section{BINARY POPULATION SYNTHESIS}
\label{sec:4}  

\subsection{BPS methods}

By performing a series of Monte Carlo BPS calculations based on the rapid binary star evolution code developed by \cite{2002MNRAS.329..897H}, we simulate the Galactic birth rate of ULXs produced from the NS+He star channel.
In each simulation, we follow a sample of $10^7$ primordial binaries until NS+He star systems are emerged. 
We assume that a ULX can be formed when the parameters of NS+He star systems are located in the initial parameter space of Figure~\ref{fig:ps}. 

Similar to \cite{2020RAA....20..161H}, the initial parameters and basic assumptions for the Monte Carlo BPS computations are shown as follows:

(1) All stars are assumed to be members of binary systems with circular orbits.

(2) The initial distribution of orbital separations (a) is supposed to be constant in $\rm log\, a$ for wide binaries, and fall off smoothly for close binaries \citep{1989ApJ...347..998E}. 

(3) The constant mass ratio ($0<q\le 1$) distribution is adopted, i.e., $n(q) = 1$.

(4) The initial mass function from \cite{1979ApJS...41..513M} is used.

(5) We adopt a constant star formation rate (SFR; $\rm5 \, M_\odot\,\mathrm{yr}^{-1}$) over the last $\rm 15\, Gyr$, in which we suppose that a primordial binary with its primary star $\rm >0.8\,M_\odot$ is produced every year \citep{1995MNRAS.272..800H,2002MNRAS.329..897H}.
A constant star formation rate of $\rm \sim 5 \, M_\odot yr^{-1}$ can be obtained based on this calibration \citep{2004A&A...419.1057W}.

It is worth noting that the NS+He star systems for producing ULXs have most likely emerged from the common-envelope (CE) evolution of giant binaries.
It is suggested that the mass-transfer process becomes dynamically unstable, leading to the formation of a common envelope (CE) when the mass ratio of the binary exceeds a critical value \citep{2020ApJS..249....9G,2023ApJ...945....7G}.
However, the CE evolution is highly uncertain.
As same as the previous studies \citep[see][]{2009ApJ...701.1540W,2010MNRAS.401.2729W}, we combine the CE ejection efficiency ($\rm \alpha_{CE}$) and the stellar structure parameter ($\lambda$) into a free parameter ($\rm \alpha_{CE}\lambda$), in which we set $\rm \alpha_{CE}\lambda = 0.5,1.0$ in this work.

\subsection{Evolutionary ways}

\begin{figure*}
\centering
\includegraphics[width=0.9\textwidth]{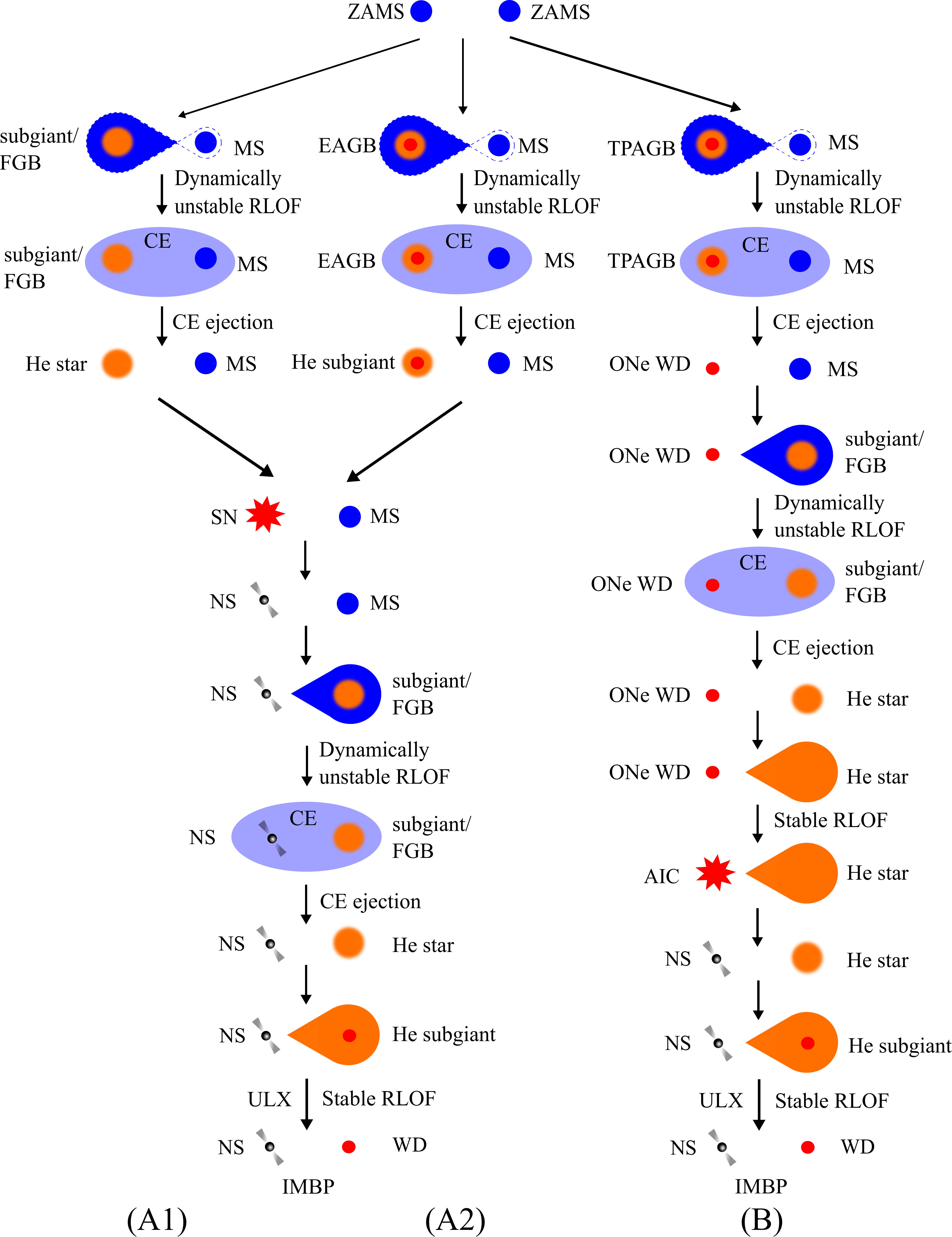}
\caption{Binary evolutionary ways to NS+He star systems that can form NS ULXs. Scenario A1 and A2 represent the CCSN channel. Scenario B stands for the AIC channel.}
   \label{fig:evolution_ways}
\end{figure*}

Figure~\ref{fig:evolution_ways} shows the three main binary evolutionary ways that can produce NS+He star systems, then evolve to NS ULXs and eventually form IMBPs, as follows:

Scenario A1: 
When the primordial primary evolves to the subgiant or first giant branch (FGB) stage, it will fill its Roche-lobe and a CE will be formed because of the dynamically unstable RLOF. 
A He star will be produced from the primordial primary after the CE ejection.
Due to the excessive mass of He star, a core-collapse supernova (CCSN) will be produced and the He star will turn into a NS.
Subsequently, a second CE will be formed due to the dynamically unstable RLOF when the primordial secondary evolves to the subgiant or FGB stage.
The NS+He star system will be produced after the second CE ejection.
When the He star evolves to the subgiant stage, it will fill its Roche-lobe and the NS+He star binary will show as the ULX during the stable mass-transfer process.
Eventually, an IMBP will be produced after the stable RLOF process.
The initial parameters of the primordial binaries in Scenario A1 are in the range of $M^{\rm i}_1 \sim 10-30\, \mathrm{M}_\odot$, $q = M_2^{\rm i}/M_1^{\rm i}\sim 0.2-0.6 $, and $P^{\rm i}_{\rm orb} \sim 100-2000\,{\rm d}$.

Scenario A2:
The primordial primary will fill its Roche-lobe when it evolves to the early asymptotic giant branch (EAGB) stage.
A CE will be formed due to the dynamically unstable RLOF and a He subgiant may be produced from the EAGB star if the CE can be ejected \citep[see][]{2002MNRAS.329..897H}.
A CCSN will be produced because of the excessive mass of He subgiant star.
The subsequent binary evolution is similar to that described in Scenario A1.
In this scenario, the initial parameters of the primordial binaries
are in the range of $M^{\rm i}_1 \sim 10-18\, \mathrm{M}_\odot$, $q \sim 0.2-1.0 $, and $P^{\rm i}_{\rm orb} \sim 1000-4000\,{\rm d}$.

Scenario B: 
When the primordial primary evolves to thermally pulsing asymptotic giant branch (TPAGB) stage, it will fill its Roche-lobe, and a CE will be formed because of the dynamically unstable RLOF. 
After the CE ejection, the TPAGB star will turn into an ONe WD.
Subsequently, when the primordial secondary evolves to subgiant or FGB stage, the second CE will be formed.
An ONe WD+He star system could be produced after the CE ejection.
The He star will fill its Roche-lobe and transfer its He-rich material to ONe WD.
The ONe WD will experience the accretion-induced collapse (AIC) via electron-capture reactions by Mg and Ne, and then evolve to a NS \citep[see e.g.][]{2008MNRAS.386..553I,2010MNRAS.402.1437H,2011MNRAS.410.1441C,2013A&A...558A..39T,wang2020formation,2023ApJ...951...91C}.
A NS+He star system will be formed after the AIC process and the subsequent binary evolution is similar to that described in Scenario A1.
In this scenario, the initial parameters of the primordial binaries
are in the range of $M^{\rm i}_1 \sim 7-8\, \mathrm{M}_\odot$, $q \sim 0.6-0.8 $, and $P^{\rm i}_{\rm orb} \sim 2000-5000\,{\rm d}$.
Compared with Scenario A1 and A2, this scenario is for the primordial binaries with long orbital periods and small primary masses.

\subsection{BPS results}

In our simulation, the Scenario A1 and A2 in Figure~\ref{fig:evolution_ways} can be collectively called the CCSN channel, because the NSs in the two scenario ways are produced from the CCSNe. 
The CCSN channel provides about $90\% $ NS ULXs with He donor stars in our simulation. 
Alternatively, Scenario B can be called AIC channel \citep[see][]{2018MNRAS.477..384L} and the NS ULXs from this channel are about $10\% $.

\begin{figure}
\centering
\includegraphics[width=0.6\columnwidth]{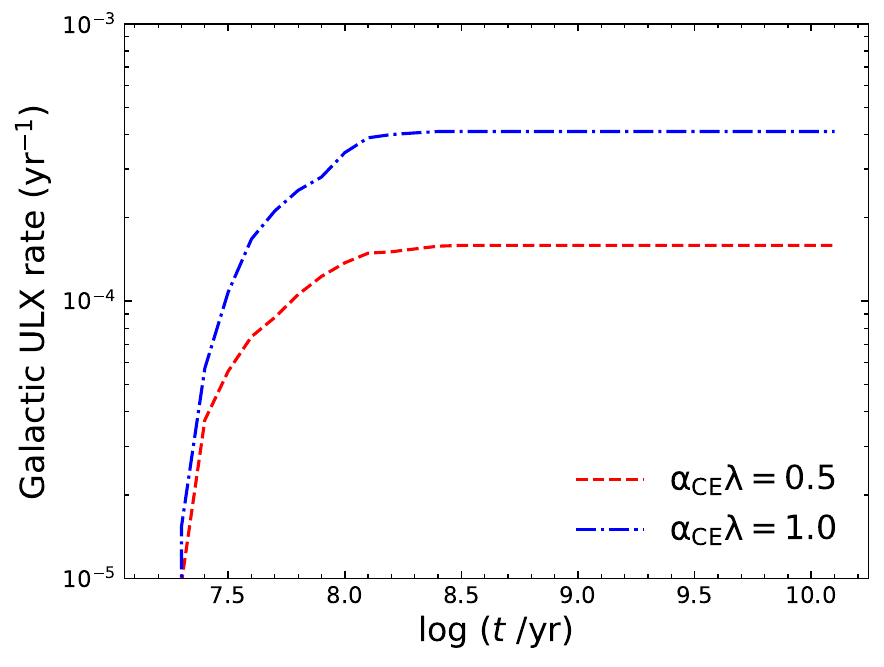}
\caption{Evolution of the Galactic ULXs rates from the NS+He star channel as a function of time.}
   \label{fig:bse}
\end{figure}

Figure~\ref{fig:bse} shows the evolution of the Galactic rate of ULXs from the NS+He star channel by adopting a constant Pop I SFR of $\rm 5 \, M_\odot\,yr^{-1}$. 
In our simulation, the most NS+He star binaries are formed with  $\rm 0.7-1.5\,M_\odot$ donor masses and $\rm 0.04-0.2\,d$ orbital periods.
The BPS calculations give the theoretical ULX rate in the Galaxy to be $\sim1.6-4.0\times10^{-4}\,{\rm yr}^{-1}$. 
If we adopt the average ULX lifetime in Table~\ref{tab:evolutionary properties} to $\rm0.278\,Myr$ and adopt the average beaming effect, there are $\sim7-20$ detectable ULXs with He donor stars in the Galaxy.
\cite{2024A&A...682A..69M} suggested that the X-ray pulses are suppressed in at least 60\% of all NS ULXs with H-rich or He-rich donors.
In this work, we adopted an average beaming factor of 0.17, indicating that about 17\% of the NS ULXs with He star donors can be detected due to the beaming effect. 
Note that the estimated number of NS ULXs with He star donors may be influenced by the selection of the simulation models, but at least providing an upper limit of its number in the Galaxy.

\cite{2019ApJ...886..118S} suggested that there are about several NS ULXs with He stars in Milky Way-like galaxy, which is slightly less than our results. 
This discrepancy may be due to several factors in our BPS simulations, as follows:
(1) SFR has large uncertainties but is important for the BPS simulations. 
\cite{2019ApJ...886..118S} adopted a SFR of $\rm 3 \, M_\odot\,yr^{-1}$, but we set the SFR to $\rm 5 \, M_\odot\,yr^{-1}$ from the calibration of \cite{2004A&A...419.1057W}.
However, \cite{2014AAS...22333604L} suggested a lower Galactic SFR of $\rm 1.66\,\pm\, 0.20 \, M_\odot\,yr^{-1}$ based on a hierarchical Bayesian statistical analysis.
(2) The mass-transfer model is also different.
\cite{2019ApJ...886..118S} employed the rotation-dependent mass-transfer model and obtained an initial parameter space containing $\rm \sim 0.6-2\, M_\odot$ He stars, which is smaller than our space.

\section{DISCUSSION}
\label{sec:5}
\subsection{Compared with previous studies}

Recently, \cite{2024A&A...682A..69M} investigated the effects of stellar age and different models on ULX populations by generating various populations at fixed burst ages.
They considered NS+He star channel for the formation of ULXs and discovered that stripped He-rich donors are prominent in ULXs around $\rm100\,Myr$, which is similar to our simulation results (see Figure~\ref{fig:bse}; the ULX rate will tend to peak after about $\rm100\,Myr$).

There is a general consensus that the majority of IMBPs evolved from IMXBs, which typically consist of a NS and a $\rm2-10\,M_\odot$ donor star \citep{1975ApJ...198L.109V}.
\cite{2013MNRAS.432L..75C} and \cite{2019MNRAS.490..752T} adopted the NS+He star channel to explain the observed IMBPs with short orbital periods ($<3\,\rm d$).
\cite{2018MNRAS.477..384L} considered the formation of IMBPs through the ONe WD+He star channel, in which the ONe WD experiences an AIC process to form a NS and the NS+He star binary eventually evolves into an IMBP.

\cite{2023ApJS..264...45F} presented a novel and general-purpose BPS code, called POSYDON. 
They also provided a binary evolution grid for $\rm 1.43 \, M_\odot$ NS$+$He star binaries (see the left panel of Figure 14 in \citealp{2023ApJS..264...45F}).
Compared to the parameter space in this work (see Figure~\ref{fig:ps}), our region shapes in which our binaries could experience stable RLOF or produce ULXs are similar to those presented in their paper, and the region in Figure~\ref{fig:ps} is slightly smaller due to the limit of the mass-transfer phase for the formation of ULXs.

\subsection{NGC 247 ULX-1}

Up to now, only one ULX with He star donor (NGC 247 ULX-1) has been identified, although it is not yet certain whether its accretor is a NS \citep[see][]{2023ApJ...947...52Z}.
\cite{2021MNRAS.507.5567D} did not find coherent pulsation with a long XMM–Newton monitoring campaign in NGC 247 ULX-1.
However, the possibility of a NS accretor in NGC 247 ULX-1 can not be completely ruled out, as its pulsation could evade detection due to the following reasons:
(1) The rotating NSs cannot be observed as pulsars if they do not have a high spin-up rate.
(2) The pulsation may be reprocessed in the stellar wind and we can only get a too low pulsar fraction in X-ray emission \citep[][]{2023ApJ...947...52Z}.
(3) The mass accretion may suppress the X-ray pulsation due to the spin alignment of the NS accretor \citep{2020MNRAS.494.3611K}.

\cite{2023ApJ...947...52Z} have estimated the orbital period of NGC 247 ULX-1 is at least $\rm2.4-21\,d$ if adopting the $\rm0.6-2\,M_\odot$ He donor star and $\rm1.4\,M_\odot$ NS accretor.
The estimated parameter of NGC 247 ULX-1 can be generally filled in our parameter space (see Figure~\ref{fig:ps} and Figure~\ref{fig:IMBP}).
In order to perform a more detailed analysis, further observations of NGC 247 ULX-1 and numerical research are required.

\subsection{Uncertainties}
In this work, our results may be influenced by some factors, e.g. the metallicity, the effect of evaporation, the accretion discs, the mass-transfer stability criteria, the NS accretion limit, etc.

It has been suggested that the formation rate of ULXs tends to increase at low metallicity 
\citep{2010MNRAS.408..234M}.
However, the nature of the accretor in ULX at different metallicities is not clear, as lower metallicities also promote the formation of BHs.
In our simulation, we aim to investigate the formation of ULXs in the Milky Way-like galaxy.
In fact, the metallicity is important for binary evolution and BPS calculations.
The initial parameter spaces would move in the direction of smaller He star mass and shorter orbital period if a lower metallicity is adopted \citep{2010A&A...515A..88W}.

The pulsar evaporation effect will evaporate the envelope of its companion with its high-energy radiation/particles if no mass-transfer interaction occurs \citep[see e.g.][]{1988Natur.334..227V,1989ApJ...336..507R,2017ApJ...851...58L,2022MNRAS.515.2725G}.
The thermal-viscous instability in the accretion discs may reduce the active X-ray time and the increase of NS mass \citep[see][]{2002ApJ...564..930L}.
The stability criteria (i.e. the critical mass ratio) during the mass-transfer process is more complex than that we adopted in this work. 
More detailed and systematic studies of critical mass ratios for dynamical timescale mass-transfer can be seen in \cite{2020ApJS..249....9G,2020ApJ...899..132G,2023ApJ...945....7G} and \cite{2024ApJS..274...11Z}.

The NS accretion limit is still uncertain.
\cite{2017Sci...355..817I} discovered that there exists an accreting NS that clearly exceeds the Eddington limit by factors up to $\sim500$ in NGC 5907 ULX.
Up to now, there are more and more observational evidence for super-Eddington accretion of NSs \citep[see a recent review][]{2017ARA&A..55..303K}.
\cite{2022MNRAS.514.1054G} considered the interaction between the NS magnetic field and the accretion disk, and adopted the critical mass-accumulation rate of \cite{2017MNRAS.470.2799C,2019A&A...626A..18C} that is about hundred times NS Eddington limit to explain the formation of mass-gap BHs.
\cite{2023MNRAS.526..854Z} suggested that the super-Eddington accretion rate in NS+He star system has to be carried out, in order to reproduce a compact binary coalescence $\rm GW\,190425$.
In this work, we still adopt the upper limit of Eddington accretion rate for accreting NSs.
We also made some tests without NS accretion limit, and found that this change only has a large effect on the final NS mass but little effect on the other final parameters.

\section{SUMMARY}
\label{sec:6}
In this work, we investigated the NS+He star channel for the formation of ULXs.
In order to produce ULXs, the NS+He systems should have He stars with initial masses of $\rm\sim0.7-2.6\,M_\odot$ and initial orbital periods of $\rm\sim 0.04-2500\,d$.
After the mass-transfer process, the ULXs with He donor stars will evolve to IMBPs, in which the WD masses is in the range of $\rm\sim0.6-1.3\,M_\odot$ and orbital periods is in the range of $\rm\sim 0.05-3200\,d$. 
Compared with the 33 observed IMBPs shown in Table~\ref{tab:IMBPs}, we found that the median parameter of the 14 IMBPs with $\rm\sim1.0\,M_\odot$ WD companions can be covered in our final parameter space.  
By using the BPS calculations, the theoretical ULXs rate in the Galaxy from the NS+He star channel is $\sim1.6-4.0\times10^{-4}\,{\rm yr}^{-1}$ and we expect to detect $\sim7-20$ NS ULXs with He star donors in the Galaxy.
In order to understand the formation of the ULXs, more observations and numerical research are needed, and large samples of observed NS ULXs with He star donors are expected.

\section*{Acknowledgements}

This study is supported by the National Natural Science Foundation of China (Nos 12225304, 12288102, 12090040/12090043, 12273105, 12273014 and 12403035), 
the National Key R\&D Program of China (Nos 2021YFA1600404, 2021YFA1600403 and 2021YFA1600400), 
the Western Light Project of CAS (No. XBZG-ZDSYS-202117), 
the science research grant from the China Manned Space Project (No. CMS-CSST-2021-A12), 
the Yunnan Revitalization Talent Support Program (Yunling Scholar Project),
the Youth Innovation Promotion Association CAS (No. 2021058), 
the Yunnan Fundamental Research Project (Nos 202401AV070006, 202201BC070003 and 202201AW070011), 
the International Centre of Supernovae, Yunnan Key Laboratory (No. 202302AN360001),
the Postdoctoral Fellowship Program of CPSF under Grant (No. GZB20240307),
the China Postdoctoral Science Foundation under Grant (Nos 2024M751375 and 2024T170393),
and the Jiangsu Funding Program for Excellent Postdoctoral Talent under Grant (No. 2024ZB705).

\section*{Data Availability}
Results will be shared on reasonable requests to corresponding author.



\bibliographystyle{mnras}
\bibliography{ULXs} 




\bsp	
\label{lastpage}
\end{document}